\documentclass[sigconf]{acmart}

\usepackage{wrapfig}
\usepackage{supertabular}
\usepackage{subfigure}

\AtBeginDocument{%
  \providecommand\BibTeX{{%
    \normalfont B\kern-0.5em{\scshape i\kern-0.25em b}\kern-0.8em\TeX}}}

\setcopyright{acmcopyright}
\copyrightyear{2018}
\acmYear{2018}
\acmDOI{10.1145/1122445.1122456}

\copyrightyear{2022} 
\acmYear{2022} 
\setcopyright{acmlicensed}\acmConference[CHI '22]{CHI Conference on Human Factors in Computing Systems}{April 29-May 5, 2022}{New Orleans, LA, USA}
\acmBooktitle{CHI Conference on Human Factors in Computing Systems (CHI '22), April 29-May 5, 2022, New Orleans, LA, USA}
\acmPrice{15.00}
\acmDOI{10.1145/3491102.3501998}
\acmISBN{978-1-4503-9157-3/22/04}

\begin{document}

\title{Model Positionality and Computational Reflexivity: Promoting Reflexivity in Data Science}



\author{Scott Allen Cambo}
\email{scottallencambo@gmail.com}
\affiliation{%
  \institution{Northwestern University}
  \streetaddress{2240 Campus Drive}
  \city{Evanston}
  \state{Illinois}
  \country{USA}
  \postcode{60208}
}

\author{Darren Gergle}
\email{dgergle@northwestern.edu}
\affiliation{%
  \institution{Northwestern University}
  \streetaddress{2240 Campus Drive}
  \city{Evanston}
  \state{Illinois}
  \country{USA}
  \postcode{60208}
}







\renewcommand{\shortauthors}{S.A. Cambo and D. Gergle}


\begin{abstract}
Data science and machine learning provide indispensable techniques for understanding phenomena at scale, but the discretionary choices made when doing this work are often not recognized. Drawing from qualitative research practices, we describe how the concepts of positionality and reflexivity can be adapted to provide a framework for understanding, discussing, and disclosing the discretionary choices and subjectivity inherent to data science work. We first introduce the concepts of model positionality and computational reflexivity that can help data scientists to reflect on and communicate the social and cultural context of a model’s development and use, the data annotators and their annotations, and the data scientists themselves. We then describe the unique challenges of adapting these concepts for data science work and offer annotator fingerprinting and position mining as promising solutions. Finally, we demonstrate these techniques in a case study of the development of classifiers for toxic commenting in online communities.
\end{abstract}







\maketitle

\section{Introduction}
Data Science and related fields like Artificial Intelligence, Machine Learning, and Statistics provide indispensable methods for extracting and understanding a variety of phenomena from large datasets. Yet, as objective as these methods aim to be, there are many undisclosed yet impactful subjective choices that data scientists make when building and deploying models \cite{passi_data_2017, passi_problem_2019, passi_trust_2018, suresh2019framework, 10.1145/3411764.3445518}.

Consider, as one example, the discretionary choices made when applying a common machine learning workflow for the development of text classifiers to detect toxic comments on social media: The data scientist must first curate a labeled dataset which requires decisions about what data is most representative of the text that the classifier will ultimately predict on. Next, they must design a labeling task that explains to a human labeler how to make decisions that will ultimately be delegated to a text classifier. The data scientist must choose which learning algorithms will be evaluated and this choice comes with the caveat that the design of each learning algorithm itself embodies a set of assumptions about the real world. Tradeoffs are made around model selection and optimization techniques that balance the need to explore a broad range of hyperparameter values with hardware limits and project timelines. When evaluating a model, the data scientist makes value decisions around how many and which kinds of errors are acceptable. Each of these decisions cascade and can lead to social and cultural misalignment with users and other stakeholders \cite{10.1145/3411764.3445518}. Text classification algorithms, like those used for content moderation, provide a good example of how models embody
a subjective perspective on what concepts like ``online harassment'' mean. 
While the data scientist may model such phenomenon as one might 
approach an objective phenomenon in physics, studies of online harassment have shown that people vary in their perception of, and reaction to, online harassment; and these variations are often influenced by identity and personal experience \cite{smith_online_2018, duggan_online_2017, duggan_online_2014}. Such concepts are highly subjective, but models
inherently adopt an objective stance.

It should come as no surprise then, that while the automated detection of toxic comments is a prolific area of research \cite{agrawal_deep_2018, al-ajlan_optimized_2018, badjatiya_deep_2017,bourgonje_automatic_2018, cheng_antisocial_2015, davidson_automated_2017, dinakar_modeling_2011, nobata_abusive_2016, wulczyn_ex_2017, yin_detection_2009, sood_automatic_2012, warner_detecting_2012}, the practical application of such technologies on real world social media platforms has had limited success. Facebook recently disclosed that only 38\% of the content flagged by users and human moderators for harassment aligned with determinations made independently by their own content filtering algorithm \cite{rosen_facebook_2018}. Such findings raise a fundamental question: How can we expect algorithms to make determinations aligned with the expectations of a broad spectrum of stakeholders when human perception of concepts like online harassment varies so considerably? More generally, the practice of machine learning and data science often come from the perspective that there is a singular, objective ground truth regarding the relationship between a target concept and the underlying data. Most methods for data exploration, learning algorithm design, and model evaluation have evolved from this central notion, and data scientists are rarely taught practices which promote acknowledgement, reflection, and discussion of their discretionary and subjective choices, though researchers have called for the need to do so (e.g., \cite{Tanweer2021Why, 10.1145/3411764.3445518, 10.1145/3442188.3445880}).

In this paper, we aim to address these deficits by drawing on qualitative concepts, methods, and practices for conducting and sharing research that incorporate subjective choices and establish norms of practice that foster critical reflection. Our contributions build upon a growing trend in the human-centered data science community \cite{10.1145/3323994.3369898, aragon_developing_2016, 10.1145/3290607.3299018, 10.1145/3406865.3418584}, and we advance three main contributions:

First, we introduce the concept of \textit{model positionality} -- the social and cultural position of a model with regard to the stakeholders with which it interfaces: the data scientist who orchestrates its development, the crowd annotators who provide the labels for the data, and the users who need to rely on the ML-based system and for whom the crowd annotators are a proxy. Together, these comprise a broad sociotechnical system that includes the model, the decision makers who affected its creation (data scientists, crowd annotators, learning algorithm) and those who interface with its deployment (those who apply it, those who are affected by it, and other peripheral stakeholders). 

Second, we introduce the concept of \textit{computational reflexivity} that aims to help data scientists and researchers apply reflexive practices to the complex sociotechnical systems and large scale data they work with. This approach, and the associated techniques we develop, serve to complement traditional reflexive methods and promote greater reflection of each stakeholder's (and model's) position with regard to the phenomena under study. We leverage the central idea that agents with similar positions will make similar decisions in similar contexts. As a first example of computational reflexivity, we develop and present a technique that extends the CrowdTruth framework \cite{mika_crowdtruth_2014, dumitrache_crowdtruth_2018} by overcoming the need to have all annotators in the system annotate all of the same data. We call the technique \textit{annotator fingerprinting}, and it can be applied to both human actors (data scientists, crowd annotators, etc.) and machine agents (learning algorithms, models, etc.). Another computationally reflexive technique we present is called \textit{position mining}, which uses annotator fingerprinting along with clustering techniques to expose commonly held positions within a large annotated dataset.

Finally, we present a case study to demonstrate the reflexive practices for data science made possible through computational reflexivity. We show how a data scientist can situate themselves within the system by annotating a sample of data and translating this to an annotator fingerprint for comparison. We demonstrate position mining to find commonly held perspectives among large sets of annotators. We also demonstrate how better reflection of the assumptions embodied in a learning algorithm can be made by varying input data and hyperparameters to see how this affects the similarity of the resulting model's fingerprint with other fingerprints.
By helping data scientists to reflect on their positionality in context with other stakeholders, as well as the models that also
embody a perspective, we believe that their discretionary decisions can be made less biased, more principled and more transparent.

\section{Background and Related Work}
\subsection{Situated Knowledge and Positionality}
\label{sec:situated_knowledge}
In 1988, Donna Haraway published an essay which sought to illuminate a tension between subjectivity and objectivity both within feminist studies and throughout 
the broader scientific community \cite{haraway_situated_1988}. Feminist scholars often wrestled with this tension in their efforts to justify the disconnect between the knowledge they produced and the knowledge produced by a generally white, male, cisgender, and privileged establishment. From Haraway's perspective, these conversations highlighted a spectrum of positions. On one end of the spectrum was what Haraway called ``radical constructivism,'' in which she critiqued that science is nothing more than a rhetorical practice aimed to persuade others and obtain `objective' power. On the other end is ``feminist empiricism,'' which yielded calls for an objective, ``successor science'' that---if done with the inclusion of women and in a more precise fashion---would result in insights that feminist scholars often uncover after deconstructing problematic knowledge produced by a western male establishment. She critiques this latter position for perpetuating ``the God trick''---presenting knowledge as objective and impartial to create the illusion of omniscience---and in doing so, hiding an often western, white, cisgendered position when a similar process applied from another position might yield critically different knowledge.

This is where Haraway finds \textit{vision} to be a helpful metaphor in understanding knowledge production. The god trick obfuscates the act of viewing and thus the viewer themselves by simply presenting an omniscient knowledge handed down from above. By centering vision as an \textit{active process in knowledge production}, it is implied that there is a viewer, that this viewer sees the world from a specific position and that their arrival at this position stems from a history which informs how and what they view. \textit{Situated Knowledge} is then presented as the idea that by acknowledging and reflecting on one's presence in the process of knowledge production, subjects can produce knowledge with greater objectivity than if they claimed to be neutral. This is because we can only claim validity if we understand and account for the conditions from which we made and understood our observation. Knowledge is most valid under these situated conditions that invites the critical dialog necessary for expanding the validity of such knowledge beyond local bounds.

Haraway emphasizes that self-presence, self-knowledge, and self-identity must be intentional and practiced in order to best answer questions like \textit{What should I be looking for?} \textit{Who should I be looking with?} or \textit{What instrument should I be using to look?} In qualitative research, this intentional reflexive practice is the development of one's \textit{positionality} or stance in relation to the social, cultural and political context of the subject. 
This gives rise to the need to be clear about which aspects of identity and one's personal experience are drawn from when producing knowledge. Making one's positionality intentional and transparent with respect to the subject of research improves validity by recognizing the social, cultural and political context of the knowledge produced.

\subsection{The Discretionary Practice of Data Science}
Researchers have recently begun to critically reflect on data science epistemology with the understanding that it is dependent on the social and political context of observation.
Passi and Jackson \cite{passi_data_2017} have described the skills needed for these tasks as \textit{data vision}, ``the ability to organize and manipulate the world with data and algorithms, while simultaneously mastering forms of discretion around \textit{why}, \textit{how}, and \textit{when} to apply and improvise around established methods and tools in the wake of empirical diversity''. If we are to consider that expertise and experience inform one's data vision, then we should also consider the social, cultural and political context of this expertise and experience.

In many ways, the concerns that feminist scholars address with concepts of situated knowledge and positionality are mirrored in recent trends within machine learning and data science. Applications of machine learning algorithms in industry have frequently presented themselves as neutral authorities with regard to the target concept -- with the implied assumption that a learning algorithm that follows a principled procedure for knowledge discovery can synthesize much more information at a faster pace and with a mechanical consistency superior to humans. As databases increase in size and more of our social interactions take place online where they are recorded at scale, machine learning is used to distill this information into a model representing a ``global perspective'' on the social phenomenon of interest. This singular, global, machine perspective is often argued to be neutral because machine learning algorithms and the models they create exist outside the social, cultural, and political contexts of humans, and thus they are safe from the unconscious biases of humans. In a sense, this is Haraway's ``god trick'' at play. By presenting algorithms and models this way, it is implied that they have greater validity.

While data science is an expansive discipline, this paper focuses on the practice of applying methods and algorithms from machine learning research. We chose this focus because it exemplifies some of the central tensions in data science work regarding the discretionary decisions and subjective interpretations that occur when automating a position with supervised machine learning techniques. While prior work has highlighted scenarios where sociotechnical challenges in machine learning work are understood by the data scientist \cite{10.1145/3411764.3445518}, they can still be limited in practice by a lack of broader organizational support \cite{passi_problem_2019} or a lack of appropriate training to address challenges \cite{10.1145/3411764.3445518}. The concepts that we introduce in section \ref{sec:computational_reflexivity} may help to bring conceptual clarity to such scenarios, and the techniques we introduce aim to produce computational and visual artifacts that can facilitate alignment and transparency of each stakeholder's
position with regard to the concept being learned by the machine learning algorithm.

\subsection{Making Discretionary Choices Salient}
Scholars have begun to consider the benefits of incorporating and reflecting on the discretionary decisions made throughout the scientific endeavor. An example can be seen in \textit{multiverse analyses} \cite{steegen_increasing_2016,10.1145/3290605.3300295} which posit that when there are multiple
reasonable ways to process a dataset or perform an analysis, then there are multiple reasonable outcomes that could be achieved. By considering all of the reasonable choices that could be made and the effects they have on the outcome, the researcher and their audience can better understand the robustness (or fragility) of the results \cite{steegen_increasing_2016}. Making multiverse analyses an explorable facet of publications can increase trust and transparency by allowing the audience to explore how the results may change under different conditions \cite{10.1145/3290605.3300295}. 
\textit{Many-model thinking} addresses a similar issue \cite{page_why_2018} and suggests that we can make better data-driven decisions when we collect a diverse group of modeling solutions, each of which errs differently and may have been constructed through different algorithmic and statistical methods. Multiverse
analysis and many-model thinking challenge the god trick by acknowledging the impact of discretionary decisions on the accuracy, fidelity and robustness of analyses.

While multiverse analysis and many-model thinking aim to increase research transparency, \textit{model cards} \cite{mitchell_model_2019} and \textit{data sheets} \cite{gebru_datasheets_2020} were introduced to increase model transparency by creating principled modes of reporting the decisions made in a data science project. Promoting a norm in data science communities of reporting in this way encourages and prioritizes transparency as well as accountability. Both model cards and data sheets provide details that give the consumer a greater understanding of the context in which the model was trained or the data were collected. This context can be viewed as a kind of situated knowledge that allows the consumers to critically reflect on their development. Gebru et al. \cite{gebru_datasheets_2020} go a step further by explicitly discouraging the automation of the creation of such datasets as this would ``run counter to our objective of encouraging dataset creators to carefully \textit{reflect on the process of creating, distributing, and maintaining a dataset.}''

Methods in data science aim to be empirically principled by drawing on principles from mathematics, computer science, and
statistics to mitigate discretionary decisions, but one could argue that discretionary decisions are inevitable. Each of the contributions described in this section provide some guidance on making these decisions more transparent for reflection by both the creator and consumer of the data, models,
and insights being produced through data science. 
In our paper, we consider how the concepts of positionality and reflexivity from qualitative research could inspire new methods that promote reflection of the social circumstances of a model's creation and application using computational methods that help us understand these relationships at the ``big data'' scale.

\section{Model Positionality and Computational Reflexivity}
Qualitative researchers in the social sciences have developed many approaches to address the challenges that subjectivity brings to scientific research. 
As mentioned in Section \ref{sec:situated_knowledge}, \textit{situated knowledge} refers to knowledge production that is contextualized with respect to a researcher's positional stance. Savin-Baden \cite{savin-baden_qualitative_2012} describes how the positional stance of a researcher results from a series of \textit{choice moments} -- times during the research process where critical decisions are made that can have an outsized influence on outcomes. 
In data science, choice moments do not only culminate in a researcher's positionality. These critical decision points also \textit{culminate in a model} which exerts influence in the sociotechnical environment. Model cards \cite{mitchell_model_2019} and data sheets \cite{gebru_datasheets_2020} 
make such choice moments transparent and salient, but it can be difficult to 
infer social influence and the social position represented in a model from 
a list of the discretionary decisions made to create it.

We contribute two points to this evolving discussion. The first is that the model(s) produced through the research process have their own positionality that should be considered in a reflexive process. As seen in prior work, data scientists often find themselves misaligned with their own models due to deficits in domain expertise or training in best practice data quality metrics \cite{10.1145/3411764.3445518}. Second, an adequate reflexive process for data science requires computational methods to scale reflexivity by an individual to large data sets and to reflect on the inherent reductions made in the modeling process. We refer to these as \textit{Model Positionality} and \textit{Computational Reflexivity},  respectively. 
In the following sections, we detail the choice moments made by a variety of stakeholders in a common data science workflow (Figure \ref{fig:ml_research_context}). 
Then we introduce the concept of \textit{model 
positionality} to describe the provenance of machine learning models designed to automate subjective decisions and their role as decision making agents in sociotechnical systems.
Finally, we introduce \textit{computational reflexivity} to describe a style of reflexivity appropriate for research with machine learning and large datasets similar to the way computational social science has complemented traditional social science methods to enable new ways of conducting research with large datasets.

\subsection{Discretionary Decisions and Choice Moments in Data Science}
\begin{figure*}
    \centering
    \includegraphics[scale=0.5]{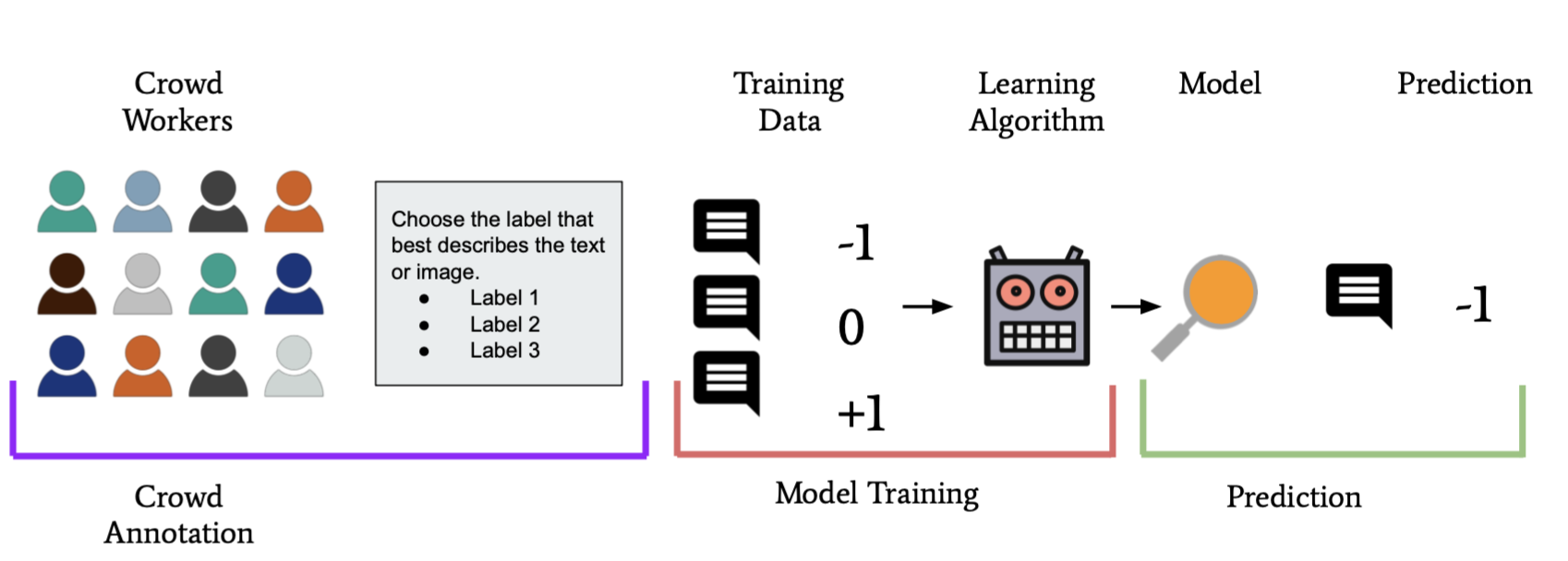}
    \caption{A common pipeline for annotating data and building models with machine learning}
    \Description{A graphical image that presents three stages of a common pipeline for annotating data. The leftmost panel is crowd annotation and has a illustration of crowdworkers and and their label choices; the middle panel illustrates training data, labels and the learning algorith; the rightmost panel illustrates the model with a magnifying glass and the prediction of minus one.} 
    \label{fig:ml_research_context}
\end{figure*}
\label{sec:choice_moments}
Figure \ref{fig:ml_research_context}, depicts a common process for 
building classification models with machine learning. At each stage, choice moments critically affect the rest of the research and model building process, but they are not always made exclusively by the data scientist; rather, decisions being made at choice moments are often a combination of those made by the data scientist, those made by crowd annotators, and those determined by computational means \cite{passi_trust_2018}. 

In the first stage of this process, a data scientist will often design an annotation task, intended to be analogous to the model's prediction task, for human labeling and human computation.
This task is then distributed to a network of crowd workers who will complete the task to
provide labels for the training dataset to be used with machine learning. While the
human decisions made in this task may not be as individually impactful as choice moments made in traditional qualitative research methodology, they are an expression
of one's personal stance with regard to the narrow slice of the research context they are
involved with. Recent research has shown that the personal stances of crowd workers can have a
significant impact on the model's final behavior \cite{10.1145/3290605.3300637}. Given 
that personal stances can be so impactful, we argue that these stances should
be considered in a reflexive process for data science.

In the model training stage, the data scientist begins to take an even more critical role in the knowledge produced. First, they must prepare the labeled dataset to be used for training.
This can typically involve validating the annotator choices through inter-annotator agreement measures like Krippendorff's alpha or Cohen's Kappa \cite{krippendorff_computing_2011, cohen_weighted_1968}. 
There is generally no consistent measure that is widely adopted in data science, nor is there a consistent threshold above which the norm in the data science community would be to reject the annotations. The choice of agreement measure and threshold are generally at the discretion of the data scientist. This validation and aggregation process creates a
singular perspective from which the machine learning algorithm will infer the target
concept and create a model of it. This is critical, because many times, the aggregation function is simply one that takes the modal response allowing even the slightest majority to prevail as the definitive source of truth for the machine learning algorithm.

At this point, the data scientist begins to focus on the machine learning 
task itself. Here, the data scientist applies assumptions regarding the 
application domain which deeply affect their exploratory data analyses as well as the models they select \cite{o2011computational, 10.1093/llc/fqn019}. These assumptions 
help the data scientist to choose the learning algorithms
that they will evaluate. Model cards and data sheets aim to facilitate reflection on these assumptions and reveal how they affect dataset
and model selection choices via documentation and disclosure of 
choices based on these assumptions \cite{mitchell_model_2019, gebru_datasheets_2020}.

\subsection{Model Positionality}
\label{sec:model_positionality}
As stated by Coghlan and Brydon-Miller, ``\textit{Positionality} 
refers to the stance or 
positioning of the researcher in relation to the social and political context of the study - the 
community, the organization or the participant group'' \cite{coghlan_positionality_2014}. Our concept of \textit{model positionality} parallels this idea by referring to the position of the \textit{model} in relation to the social, cultural and political context, and includes both its provenance as well as its deployment since the two are so deeply intertwined via the goal of model generalization. 

To fully capture model positionality, we need a way to characterize a model’s behavior in context with the behaviors of humans executing the same task or a reasonable proxy. By positioning the model in a space among appropriate human counterparts, we not only learn which perspectives the model is representating and automating, but also \textit{which perspectives the model has overlooked or may be systematically at odds with}. While models can incorporate data from a broad variety of perspectives and be deployed in a variety of situations and contexts, we should not confuse virtual omnipresence with omniscience. Model positionality is a characterization of the position from which a model ``views'' the world. As such, model positionality should also acknowledge its partial perspective of the world and invite reflection on what is automated in this partial perspective.

\subsection{Computational Reflexivity}
\label{sec:computational_reflexivity}
Reflexive practices are intended to help researchers to consider how their position may have detrimental or underappreciated influence on a study.
There is perhaps no field in better need of such reflection than data science. Data science applies knowledge of statistics and computation to a broad range domains from agriculture to automated driving to content moderation to wine selection. Data scientists also have eclectic educational backgrounds which means methodological and epistemological positionalities
based on these backgrounds are likely to vary 
considerably \cite{lindner_where_2019}.
Recent work has argued for the incorporation of qualitative thinking, particularly reflexivity. Tanweer and colleagues \cite{Tanweer2021Why} describe 
qualitative `sensibilities` and connect them to data science work while providing three examples of reflexive techniques that 
data scientists can use. Miceli and colleagues \cite{10.1145/3442188.3445880} argue that reflexive practices may help address some of the more salient issues with documenting datasets for computer vision. While many reflexive practices for qualitative research can be adopted in data science (e.g. reflexive journaling, brain dumps, situational mapping, toolkit critiques), there are 
some distinctions that merit the use of computational methods.

The first major distinction is that knowledge of the phenomenon is not only understood through direct observation or 
interaction with those who have personal experience as might be done with ethnographic methods. Typically, this knowledge is produced 
through brute force pattern recognition and statistical analyses that often require thousands of explicit examples. This 
creates an issue of scale not easily addressed with most traditional approaches. 
Given this, we suggest that data science will need 
to innovate and incorporate new computational methods designed to facilitate reflexive practices that can address the challenge
of scale. 

The second issue is that, while the data scientist is often responsible for orchestrating the entire knowledge production process, their positional
stance is not the only factor influencing the final output. Crowd annotators, both individually and collectively, exert positional influence through their interpretation of labeling tasks and labeling choices that ultimately influence outcomes \cite{10.1145/3290605.3300637}. Furthermore, as observed in \cite{passi_problem_2019}, data 
scientists themselves must also negotiate their conceptualization of the problem with broader members of the organization in which they work. 
All of these
positional stances and influences comprise \textit{a system of knowledge production} that can be even broader than the sociotechnical environment wherein the model will be embedded. Each of these
various sources can influence other components of the process and should be understood by a reflexive practice that adequately recognizes and mitigates harmful social biases. 

Together, these various distinctions and challenges merit a computational approach to reflexivity. As such, we define \textit{computational reflexivity} as a set of computational methods and ways of thinking that facilitate reflexivity in data science. Stated another way: Computational
reflexivity is to reflexivity as computational social science is to social science. Computational
reflexivity is not intended as a replacement for reflexivity, but rather as a collection of methods that enable 
reflexivity to occur in new contexts. In the following sections, we present 
several methods for consideration in computational reflexivity as well as the concept of model positionality to describe the culmination of positional stances and subjective knowledge encoded in a computational model.

\subsection{Practical Challenges: The Limitations
of a Demographic Lens for Model Positionality}
Current research in model bias is often done through a demographic lens which uses a factor such as gender to split the data into groups that can be modeled or evaluated individually. This approach has some practical strengths. One is that many demographic traits are often part of annotator profiles and can be leveraged as a way to select specific demographic groups when deploying an annotation task. 
This allows the data scientist to create a more equitable annotator pool.
Another potential strength is that demographic traits are often the subject of social science research allowing the data scientist to assess the risk of model bias by connecting the demographic groups in their training data to more rigorous demographic-based studies of behavior. For example, data scientists concerned about aspects of gender-based predictive bias in their model can draw on research which discusses how gender bias is expressed in textual language and the effect it has in online communities.

However, there are several assumptions made when working with demographics that can fail when applied to model positionality. The first is that when data scientists split annotators into groups based on a demographic trait, there is an assumption that members within each group will exhibit similar judgements. However, prior work shows
that this is not always true. Take, for example, the women who annotated a toxic content dataset. They exhibited lower inter-annotator agreement than the men \cite{binns_like_2017}. When there is low agreement within a group, it is difficult to make claims about the group's collective position, which then makes it difficult to make claims about how well a model represents that group. One might assume that low agreement within a demographic group might be remedied by further splitting the group based on additional traits. For example, we might split the group of women annotators by race to understand if any of the new subgroups, e.g. black women and white women, exhibit a higher inter-rater agreement. What makes this assumption difficult to operationalize stems from both theoretical and practical concerns. There is no guarantee that a dataset will include the demographic information needed to account for the low within-group agreement. Furthermore, judgements with regard to subjective concepts such as toxic comments are not always consistent with demographic categories or a notion of identity that can be easily surveyed. This means that we may need to consider looking for positions that are not associated with traditional demographic traits. While investigating model bias by demographics can sometimes be a useful approach, understanding the subjectivity of a socially constructed concept and a model's positionality with regard to that concept, may be better understood through methods which can distill common perspectives from annotated data while being agnostic to demographic traits. 

A common assumption is that if a demographically imbalanced dataset yields imbalanced model predictions, then a demographically balanced dataset must yield balanced model predictions. This assumption can fail, because learning algorithms have an inductive bias that can unintentionally favor one group or another regardless of representation in the dataset. Prior work on toxic comment classification shows that even when some models are trained exclusively on labels contributed by women, the predictions align more closely with contributions by men \cite{binns_like_2017}. This bias may not be intentional, but its propagation from model training through model validation and selection can result from it being treated as a complete representation of one's perspective, social position, or identity.
By creating tools and methods that allows us to conceive of annotator positions as a function of the annotation judgements they make and the contexts in which they make them, we can more critically investigate these assumptions to potentially reduce harmful biases in machine learning systems.


\section{Annotator Fingerprints: A Data Structure for Computational Reflexivity}
\label{sec:annotator_fingerprinting}
Reflexivity requires that we systematically consider the social and cultural context of the observer's position in a principled way. To better understand the social and cultural context of a machine learning classifier, the data scientist should be able to systematically compare a model's prediction behaviors with the annotation choices of relevant human stakeholders, particularly those of the data scientist and the crowd annotators\footnote{In scenarios where such data are available, we can consider user behaviors as well.}. We introduce \textit{Annotator Fingerprints} as a data structure that can be used as a digital representation of the annotation behaviors of any annotating agent. In this sense, we use the term annotator broadly to include not only human annotators, but also machine classifiers. We define \textit{Annotator Fingerprinting} as the process of developing a pragmatic representation of annotators with regard to the items they label and the labels they attribute to them. While similar to the concept of \textit{task fingerprinting} described in \cite{rzeszotarski_instrumenting_2011}, this process aims to be specific to the task and to make valid comparisons between annotators even when annotators have annotated very few items in common. 

\begin{figure}[ht]
    \centering
    \includegraphics[scale=0.7]{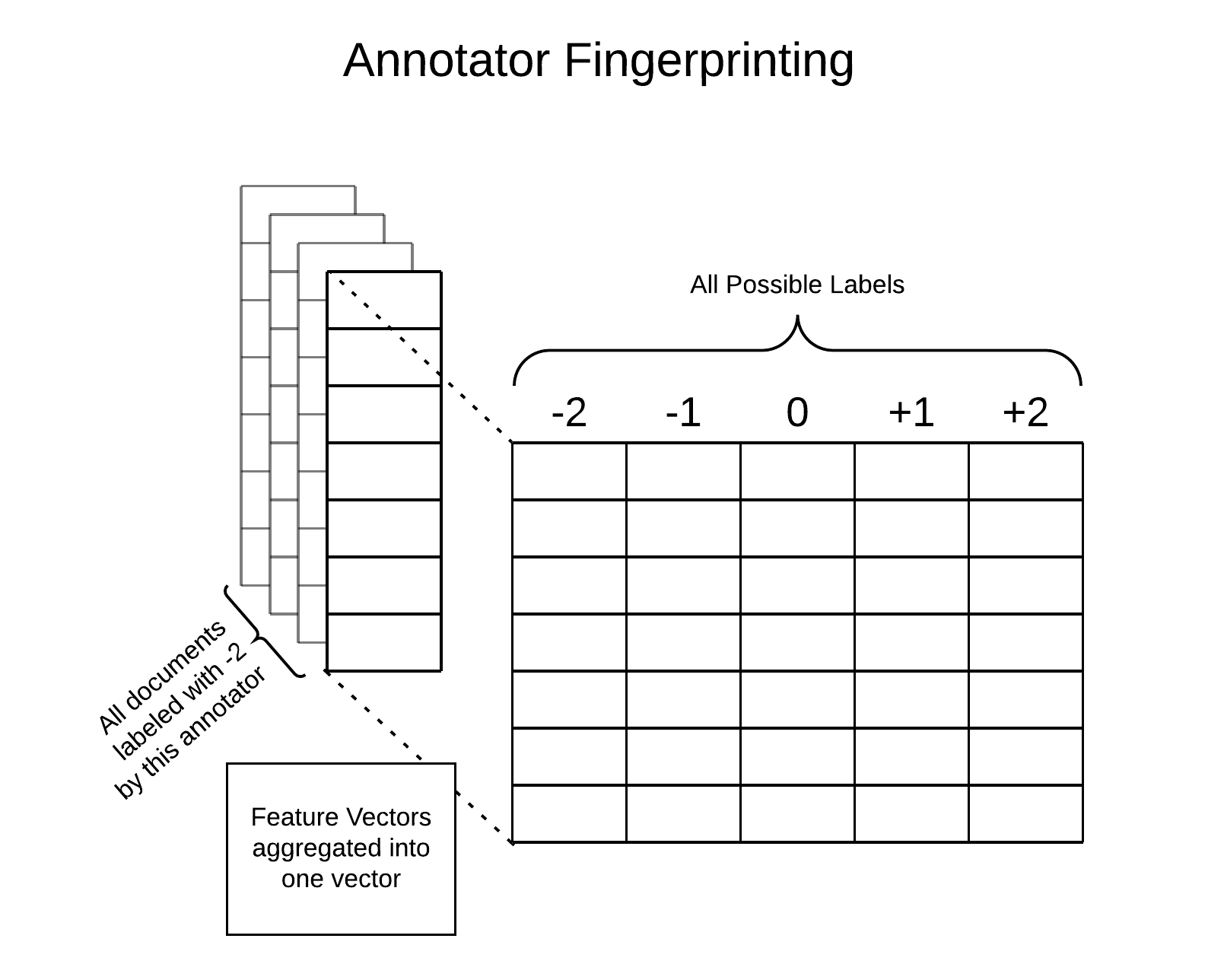}
    \caption{Illustration of the \textit{annotator fingerprint} 
    for a single worker in which all of the fingerprinting features extracted from documents that were given a particular label by the annotator are aggregated together to create a single vector. The fingerprinting vectors for each label type become the columns of the annotator fingerprint matrix.}
    \Description{This is a black and white illustration of the annotator fingerprinting vectors. On the left is a set of columns illustrating all of the documents with a label of minus two. On the right is the main matrix with column headers from minus two to plus two - indicating the possible labels.} 
    \label{fig:annotator_fingerprinting}
\end{figure}

The Annotator Fingerprinting technique is motivated by and expands upon the CrowdTruth framework \cite{mika_crowdtruth_2014, aroyo_three_2014, aroyo_truth_2015}. CrowdTruth is an ongoing research program in crowd annotation that aims to develop a richer understanding of the dynamics among annotators, annotations, and the content being annotated \cite{aroyo_harnessing_2013, aroyo_three_2014, aroyo_truth_2015, dumitrache_crowdsourcing_2018, dumitrache_crowdtruth_2018, soberon_measuring_2013}. A core tenet of the framework is that annotator reliability metrics should be able to preserve and represent disagreement, as it is more often a signal of semantic ambiguity than annotator error \cite{soberon_measuring_2013, aroyo_harnessing_2013}. 

At the core of CrowdTruth is the `triangle of disagreement' which represents the relationships between the annotators, the labels, and the items being labeled such that metrics can be designed by studying the distribution of values across one axis to learn about the other two \cite{aroyo_three_2014}. 
The result of a single worker annotating a single item is represented by a binary vector within each row of the tables representing each worker. This is referred to as the \textit{worker-unit vector}, $V_{w,u}$ where $w$ represents the specific worker and $u$ represents the specific item, or \textit{media unit} as they are referred to in the CrowdTruth framework \cite{soberon_measuring_2013}. 
The \textit{Media Unit vector}, $V_u$, can be computed by summing all worker-unit vectors for a particular media unit: $V_u=\sum_w V_{w,u}$.  This representation can be used for measuring worker-to-worker agreement by doing pairwise comparisons between each pair of worker vectors.

The triangle of disagreement is a powerful way to debug an annotation task, but when there is little per-item overlap
among annotators---as is the case with the Wikipedia Toxic Comments dataset that we'll describe in Section 
\ref{sec:wiki_dataset}---the authors note that the approach is less valid \cite{aroyo_three_2014, dumitrache_crowdtruth_2018}. 
To resolve this, we use topic modeling to decompose individual media units into more fundamental 
and comparable parts. This effectively sacrifices the specificity we get from measuring 
annotation behaviors by specific media units in favor of better validity with
respect to annotator comparisons. In essence, \textit{Annotator Fingerprinting} creates a matrix representation of the annotator (see Figure \ref{fig:annotator_fingerprinting}) that can be used to derive agreement metrics and analysis methods like those proposed by the CrowdTruth framework, but deal with the annotator sparsity problem by developing a ``fingerprint'' of annotation behavior for each annotator.

The format of the annotation fingerprint is intended 
to be compatible with many of the measuring concepts described in \cite{soberon_measuring_2013, aroyo_three_2014, aroyo_crowdsourcing_2019} except that we represent the items being labeled more broadly than specific media units to address sparsity in annotator overlap. To address this sparsity and make annotator fingerprints represent annotators in a way that permits effective comparison of their annotation behaviors, we make an assumption that \textit{annotators with similar positions will label similar data with similar labels}. To represent ``similar data'', we apply topic modeling, which allows us to identify recurring patterns of co-occurring words in the corpus. Specifically, we use Latent Dirichlet Allocation (LDA) \cite{blei-ng-jordan-LDA-2003} which can be thought of as a pseudocount representing the number of times a word, $w$, was assigned to topic, $t$. Sorting these words for the highest count for a topic yields a collection of words that typically describe that topic well. 

By decomposing each document of a corpus into a set of core components to be measured, we can compare annotators by their affinity for attributing certain labels to documents with each of these components. To create the annotator fingerprint, we then group all documents labeled by an annotator by the label they attributed and aggregate these topic vectors\footnote{In topic modeling, a decision about the number of topics must be made prior to learning the model. We evaluated models from 2 to 20 topics and calculated their training perplexity and perplexity on a held out test set. A model with 13 topics was chosen based on the best training and testing perplexity scores. Further details on the LDA topic modeling procedures can be found in \cite{cambo_2020}.}. The result is a representation illustrated in Figure \ref{fig:annotator_fingerprinting} where each row represents a topic and each column represents the annotation label. 

\section{Methods for Using Annotator Fingerprints}
\label{sec:annotator_fingerprint_methods}
In this section, we present several methods designed to
use annotator fingerprints for computational reflexivity.
This is not intended as an exhaustive set of uses, but rather serves to provide examples of some of the reflexive practices they can enable.

\subsection{Annotator Fingerprint Similarity for Pairwise Agreement}
\label{sec:af_similarity}
For the case study that we present in Section \ref{sec:case_study}, each worker's annotation fingerprint is transformed into a 1-dimensional vector in row major order (by topic) of the original matrix. These vectors are then compared using cosine similarity. This approach assumes that the relationship between values in a nearby position in the annotator fingerprint have little impact on assessing similarity. We know that this assumption is, at least, somewhat naïve because these labels are ordinal, meaning that there is some interdependence with nearby values. Other approaches can be devised by drawing inspiration from image similarity in which matrices are compared for structural similarity \cite{wang_context_2004-1} or from other similarity measures which are based in measuring divergence from the average worker-label vector \cite{dumitrache_crowdsourcing_2018}.

\subsection{Position Mining}
\label{sec:position_mining}
In crowd annotation work, divergent labels are often interpreted as ``noise'' or an indication of poor work quality when it would be equally valid to consider this an indication that there are multiple valid perspectives on the concept \cite{mika_crowdtruth_2014}. Currently, there are gaps in existing methodology that make it difficult to consider this alternative and prevent us from answering questions like:
\begin{itemize}
    \item \textit{How do we know when divergent labels indicate that there are multiple perspectives at play?}
    \item \textit{Which annotators are likely to have similar or different perspectives?}
    \item \textit{How divergent are the groups of annotators who have similar perspectives?}
    \item \textit{In what contexts do annotators diverge most or least?}
\end{itemize}
We propose an approach to answering these questions that we call \textit{position mining}. Position mining is the use of statistical and machine learning techniques, such as unsupervised clustering, to identify groups of annotators who apply similar annotations to similar items in a corpus or dataset. 

Position mining enables the data scientist to make deliberate choices about the social impact of their models by identifying salient annotator positions within the annotated data. For example, if a mined position seems to systematically label content with markers of African-American English (AAE) as toxic, as has been found in prior work \cite{sap_risk_2019}, we might choose to down-weight or outright ignore the label contributions of those within that positional group in order to prevent downstream consequences like systematically silencing an entire culture through automated content moderation. These deliberate choices can be made transparent by documenting them in model cards \cite{mitchell_model_2019} or writing a positionality statement for stakeholders such as the engineers who are responsible for incorporating the model into a fully functional system.  

In the case study below (section \ref{sec:case_study}), we use the annotator fingerprint pairwise agreement, described in section \ref{sec:af_similarity}, as the similarity function for an unsupervised clustering approach. While we use the DBSCAN clustering algorithm \cite{schubert_dbscan_2017}, we found that introducing an intermediate
dimensionality reduction step with UMAP \cite{mcinnes_umap_2018} had a 
significantly positive effect on cluster validity as measured by silhoutte scores ($.29$ to $.37$). We use a density-based clustering algorithm, because it requires no assumptions about the number of clusters we expect to find and 
it favors more cohesive clusters over the ability to assign each data point to its
most likely cluster. The result is that many data points are not assigned to a 
cluster, but the clusters that are derived represent a more salient
and consistent position in the data.

\subsection{Visualizing Annotator Fingerprints}
\label{sec:af_viz}
When we create annotator fingerprints of thousands of crowd workers, the data
scientist, and many versions of a model trained under different conditions, 
it can be difficult to understand which are most aligned and where divergences
occurred in the process of building the model. To address this issue, we can
use visual analysis methods which reduce each fingerprint to a 2-dimensional 
coordinate that can be graphed in a scatter plot. For our case study, 
we use the UMAP dimensionality reduction method to collapse the 
flattened annotator fingerprint into a 2-dimensional representation \cite{mcinnes_umap_2018}.
This visual analysis method creates an entry point for further investigation that allows the data scientist to query the nearest neighbors of a 
fingerprint in this space to see which annotators (human or machine) have similar fingerprints and thus are most aligned positionally. When the data scientist
observes their own annotator fingerprint among the annotator fingerprints of the 
model and crowd workers, they are seeing a representation of their positionality within the broader research context, satisfying a primary goal of reflexivity.

\section{Case Study: Toxic Content Modeling}
\label{sec:case_study}
We now demonstrate a case study of computational reflexivity in the context of toxic content classification. This reflexive practice was done jointly as a group, representing the reflection of the authors of this paper. We report on a first-person account and begin by detailing the ways in which computational reflexivity can be enacted in a data science setting. We then describe the dataset used in the case and discuss why it was chosen. We perform position mining and present the results in the context of the dataset. Finally, we conclude with a demonstration of the reflexive use of annotator fingerprinting for understanding the relationships among analyst, annotator and model perspectives. 

\subsection{A Computationally Reflexive Approach to Modeling Toxic Content}
A common epistemological thread among many data scientists is an appreciation for math, statistics, and computer science -- fields which have traditionally focused on using quantitative insight to construct ``objective'' knowledge. However, practicing data scientists often work with subjective, socially constructed and highly contested concepts. In these cases, a reflexive approach can help data scientists to wrestle with such facets of subjectivity. This case study illustrates the ways in which a data scientist can practice reflexively in their research by: 
\begin{enumerate}
    \item providing a way to reflect more deeply on the content, particularly in places where they feel
    unable to determine toxicity, and understand how and why it may be challenging.
    \item providing a way to understand their personal position with respect to the broader pool of annotators.
    \item providing a way to understand the position of a trained model in the context of this richer, more contextualized understanding of the items, annotators and one's own position.
\end{enumerate}

\subsubsection{Identifying and Acknowledging Subjectivity}
A core demand of data science is to balance the inherent complexity and uncertainty of the real world with the need to make simple and concrete formalized assumptions when building models and designing algorithms. When data scientists use machine learning with crowd annotated datasets to learn a socially constructed concept like ``toxicity,'' they often attempt to distill multiple different perspectives into a single one. However, in order to take a reflexive approach the data scientist needs to incorporate an understanding of multiple perspectives toward the target concept during the exploratory data analysis stage. For labeled data, inter-annotator agreement measures can help to surface ambiguous labels, but they do not enable the deeper analysis needed to distinguish divergent labels caused by an ambiguous annotation task design, too few workers, poor work quality, or truly divergent perspectives toward the concept. To address this, we have developed position mining and annotator fingerprints as methods to enable this deeper, multi-perspective analysis.

\subsubsection{Data Scientist Positionality}
After identifying and acknowledging the subjectivity of the concept being modeled, the data scientist can engage in a reflexive process similar to that of qualitative researchers. Once a data scientist understands their position with respect to the target concept situated within the context in which they are working, they can more intentionally and transparently make the discretionary decisions needed. We will demonstrate how a data scientist can annotate data from a labeled dataset to ``locate'' themselves with respect to other annotators. This process provides an opportunity for reflection and exploratory data analysis regarding a variety of common perspectives that account for the divergent annotations in the data.

\subsubsection{Representing Model Positionality}
As argued in our discussion of model positionality, models often automate a single perspective toward the target concept even when the labeled training data represent a variety of perspectives. Once we acknowledge that the target concept is subjective, it is important to compare the position of a model we are considering with the variety of positions that are held by the users who will be affected by the model -- in essence, clarifying whose position is ultimately privileged by the model. While it is common to believe that a model's prediction behaviors and predictive biases are solely dictated by the training data and thus \textit{only} the training data should be manipulated to mitigate this predictive bias, we also need to attend to the role of inductive bias in the learning algorithm. Using the annotator fingerprinting process, we can build \textit{model fingerprints} to locate a model with respect to the data scientist and the annotators of the training data. These model fingerprints are a representation of model positionality in the same way that annotator fingerprints are a represenetation of annotator positionality. We'll show that the data scientist can use model fingerprints with different subsets of the training data and its annotations to better understand the data's effects on the final model's predictions, as well as which groups of people and thus which perspectives will benefit from the prediction tendencies of the model.

\subsection{Wikipedia Toxic Comments Classification Dataset}
\label{sec:wiki_dataset}
We use the Wikipedia (WP) Toxic Comment Classification Dataset \cite{wulczyn_wikipedia_2017, wulczyn_wikipedia_2017-1} which is organized in three parts: the demographic information for each annotator (n = 3,591), the Wikipedia talk page comments being annotated (n = 159,686), and the individual annotations for each comment (n = 1,598,289). 
Demographic information includes age, gender\footnote{It's important to note that annotators were only given the option of choosing \textit{male}, \textit{female}, or \textit{other}. Only one annotator chose the gender identity of \textit{other}.}, educational background, and whether English was the annotator's first language. Importantly, the data set includes unique identifiers for each annotator that can be used to relate them to their specific annotations. 

\begin{figure*}[t]
    \includegraphics[width=\textwidth]{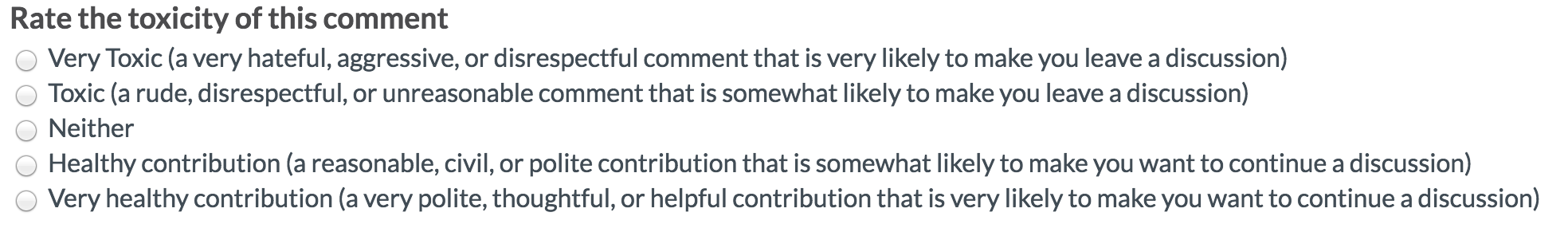}
    \caption{The annotation questionnaire shared with crowd annotators in labeling the Wikipedia Toxic Comment Dataset.}
    \Description{This is a screenshot of the toxicity rating scale used by the annotators. It includes a title saying rate the toxicity of this comment and then five stacked radio buttons that range from very toxic to very healthy (top to bottom).} \label{fig:toxicity_question}
\end{figure*}

The data were initially collected using the CrowdFlower\footnote{CrowdFlower is now known as Figure Eight.} platform where each crowdworker (i.e., annotator) was shown a random comment drawn from a subset of Wikipedia Talk Pages (i.e., discussion pages for editors). Annotators were asked to ``Rate the toxicity'' of a comment ranging from ``(-2) Very Toxic'' to a ``(+2) Very healthy contribution''. See figure \ref{fig:toxicity_question} to see the annotation interface and refer to \cite{wulczyn_ex_2017} for more details on the annotation process. Each comment was labeled by ten annotators and we make use of the unaggregated label set. Somewhat surprisingly---given the highly influential nature of the dataset---item reliability is extremely low (Krippendorff's $\alpha=0.45$). For context, Krippendorff suggests that it is ``customary to require $\geqslant .800$. Where tentative conclusions are still acceptable, $\alpha \geqslant .667$ is the lowest conceivable limit'' \cite{krippendorff2018content}. This suggests marked variability across the annotators and suggests that there may be numerous different perspectives with respect to the target concept of toxicity. 


We chose the WP Toxic Comment dataset for numerous reasons. First, it is highly influential in both academia and industry. Machine learning researchers often use the dataset for training novel algorithms \cite{agrawal_deep_2018, binns_like_2017, elsherief_peer_2018, pavlopoulos_deeper_2017, pavlopoulos_deep_2017, dixon_measuring_2018}. Jigsaw, a Google company, used the dataset to build models of toxic behavior for their Perspective API \cite{dixon_perspective_2019}. Furthermore, Jigsaw has worked with Kaggle to run two separate competitions using the dataset. One in which kagglers compete to build models optimized for maximum accuracy \cite{knapp_kaggle_nodate} and the second, to optimize for both maximum accuracy and minimum ``unintended bias'' \cite{noauthor_jigsaw_2019}. Second, the dataset is typical of those used for toxic comment classification in that it collects labels for comments without any surrounding conversational or topical context. While this has been shown to be insufficient for making an accurate determination \cite{sood_automatic_2012}, it is nonetheless a standard approach used in many systems \cite{agrawal_deep_2018, elsherief_peer_2018, pavlopoulos_deeper_2017, department_of_informatics_and_systems_engineering_technical_university_of_moldova_inter-annotator_2017, dixon_measuring_2018}. Third, we have basic demographic information for each annotator which allows us to relate differences in perspectives among annotators to existing research such as research studying the role of identity in experience with online toxicity \cite{duggan_online_2017, duggan_online_2014, smith_online_2018}. Finally, the unique IDs for each annotator in this dataset are critical for our research. Without them, it would be impossible to study annotator positions, because we would not know which labels were provided by the same annotators \footnote{Aggregated annotations only allow us to study and build models based on an aggregated perspective and interpretation of the annotation task. This can lead to problems in which the final model fails to represent large portions of a community simply because they were not represented by the majority of annotators.}. 

While both men and women cover most of the items in the dataset (99.97\% and 99.28\% respectively), the average number of men annotating a given comment is far higher (M = 5.57 and M = 2.97, respectively). This means that while coverage is similar, men may still consistently ``outvote'' women when it comes to content that may be controversial with respect to gender. The age groupings show a far more exacerbated bias with 85\% of annotators being between the ages of 18 and 45 years old. The intersections of these demographic categories (e.g., annotators who both identify as \textit{female} and are \textit{45 to 60 years old}) are much smaller, presenting a challenge for
those attempting to exhaustively study the relationship between intersectional identities,
human annotation, and machine learning model prediction with these data. 

\subsection{Position Mining Results}
To understand whether there may be multiple major perspectives with regard to the target concept of
toxicity, we used \textit{position mining} to understand annotators as the nearest proxy we may have
to the end-users of the system this model would be deployed in. The goal of position mining is to
reveal groups of annotators who apply similar annotations to similar comments in the dataset. 
By making use of annotator fingerprinting and unsupervised clustering methods to perform position mining,
we found two primary clusters. The positional clusters derived from this position mining method exhibited
much greater cohesion and separation as measured by silhoutte scores \cite{rousseeuw_silhouettes:_1987}
than clusters based purely on demographic factors ($0.37$ for positional clusters while demographic
groupings all yielded less than $0.02$). This is worth noting given that many approaches to investigating
algorithmic bias are done through the lens of demographics (for full details see \cite{cambo_2020}). Of the positional clusters, Cluster 0 consisted of $1,730$ individuals
and Cluster 1 consisted of $900$ individuals (the clusters are visualized in Figure \ref{fig:ann_fp_w_ds}).

\label{sec:position_mining_study2_results}

\begin{table}[t!]
    \begin{tabular}{llr}
    \hline
    \textbf{Category} & \textbf{D} & \textbf{Holmes-Bonferroni Adj. P} \\
    \hline
    \textbf{negative\_emotion} &  0.187508 &           7.222302e-09 \\
    \textbf{death           } &  0.179592 &           7.540909e-09 \\
    \textbf{children        } &  0.179230 &           2.947157e-09 \\
    \textbf{business        } &  0.176636 &           3.396047e-09 \\
    \textbf{power           } &  0.175691 &           4.792485e-09 \\
    \textbf{ridicule        } &  0.174921 &           4.102396e-09 \\
    \textbf{speaking        } &  0.173894 &           3.165459e-09 \\
    \textbf{shape\_and\_size  } &  0.171728 &           4.837695e-09 \\
    \textbf{swearing\_terms  } &  0.171047 &           3.052410e-09 \\
    \textbf{emotional       } &  0.166960 &           7.024445e-09 \\
    \textbf{listen          } &  0.165517 &           3.146039e-09 \\
    \textbf{healing         } &  0.164634 &           4.459110e-09 \\
    \textbf{giving          } &  0.164079 &           2.713262e-09 \\
    \textbf{optimism        } &  0.163336 &           3.770597e-09 \\
    \textbf{injury          } &  0.162214 &           4.538029e-09 \\
    \textbf{positive\_emotion} &  0.161338 &           3.944618e-09 \\
    \textbf{hearing         } &  0.160655 &           9.495774e-09 \\
    \textbf{violence        } &  0.160393 &           2.771926e-09 \\
    \textbf{toy             } &  0.159801 &           4.135478e-09 \\
    \textbf{internet        } &  0.159574 &           2.050861e-08 \\
    \textbf{movement        } &  0.157212 &           3.975195e-09 \\
    \textbf{reading         } &  0.156064 &           3.586029e-09 \\
    \textbf{driving         } &  0.155480 &           4.203270e-09 \\
    \textbf{heroic          } &  0.153959 &           3.015682e-08 \\
    \textbf{aggression      } &  0.153636 &           1.349346e-08 \\
    \bottomrule
    \label{table:ks_test_empath_categories}
    \end{tabular}
\caption{The results of 2-sample Kolmogorov-Smirnov tests which reject the null hypothesis with a Holmes-Bonferroni adjusted p-value. Displayed are the top 25 Empath categories by $D$ value which passed the test.}
\label{table:ks_test_empath_categories}
\end{table}

We then chose to further investigate the clusters to more meaningfully reflect upon a complex activity like annotating toxic behavior. To do so, we applied the \textit{Empath} Python package \cite{fast_empath:_2016} to quantify 210 different lexical categories and interrogate the types of language that best account for different annotation patterns among the cluster annotators. 
For each worker, we sum the word counts for each lexical category among the documents that the worker judged as ``moderately'' or ``severely toxic''. For each lexical category and cluster, we then create a new vector in which each element represents an annotator and the value represents the number of times a document with a word associated with this lexical category was deemed toxic. This captures patterns in lexical categories associated with toxicity as rated by a given annotator. We can then compare how these assessments differ between the set of annotators residing in cluster 0 versus the set of annotators residing in cluster 1.

For each Empath linguistic dimension, we statistically compare the distribution of counts in clusters by applying a two-sample, asymptotic Kolmogorov-Smirnov (KS) test. The KS test is a non-parametric, distribution-free test that is robust to outliers \cite{jr_kolmogorov-smirnov_1951} and allows us to compare the two sample distributions and determine whether or not they are likely to come from the same probability distribution\footnote{The null hypothesis ($H_0$) is that the two sample distributions are drawn from the same probability distribution. Rejection of the null hypothesis implies that the two groups were likely sampled from populations with different distributions – in other words, the annotators from each cluster appear to view toxicity differently relative to the given lexical category.}. We calculate the Kolmogorov-Smirnov $D$ test statistic for each linguistic dimension, sort linguistic dimensions by D value to reveal the lexical categories where there is the largest difference between the clusters, and perform the KS test to determine whether the differences are statistically significant.\footnote{Due to the large number of comparisons being performed (one for each of the 210 lexical categories) there is an inflated chance of a Type I error (i.e., a false positive). To adjust for this, we apply Holm’s Sequential Bonferroni Procedure (i.e., the Holm-Bonferroni method) and only discuss results that pass this more stringent significance requirement.}

Table \ref{table:ks_test_empath_categories} shows the lexical categories which passed this test and are among the 25 highest $D$ values from the test. Among these top distinguishing lexical categories, four of them are explicitly related to affective states (\textit{negative\_emotion}, \textit{emotional},  \textit{positive\_emotion}, and \textit{aggression}) as well as \textit{swearing\_terms} and \textit{ridicule}, which are more implicitly related to affective state.
Comments which have high word count in these categories can represent a heated debate or conflict in deciding what should or should not be on a particular Wikipedia article. One potential reason for the divergence between these clusters may be differences in a sense of decorum. Further investigation of the comments labeled as far more toxic by the larger cluster (i.e., Cluster 0) than the smaller one (i.e., 
Cluster 1), reveals bits of conversation in which the editor uses profanities and graphic language to express what they believe is best for the article. While we cannot say with certainty that members of the larger cluster value decorum more than the crude, but cathartic rhetoric that seems to be favored by the smaller cluster, comments like the following (rated as ``-2.0 (Very toxic)'' by Cluster 0 and ``1.0 (Healthy contribution)'' by Cluster 1) characterizes a common pattern and seems to suggest as much:

\begin{quote}
\textit{``No. you shut up, famousdog. You ARE a racist you vehemently refuse to accept the validity of a philosophical paradigm which is alien to the Western Scientific paradigm youre attached to. You will never have an accurate understanding of acupuncture if you are too xenophobic to adopt its endogenous native context, which is inherently Chinese, and seek to criticise it from a purely Western point of view.''}   
\end{quote}

Another interesting insight is that while many of the cases of words related to \textit{death}, \textit{violence}, or \textit{injury} stem from editors speaking about committing violence, others come about from language used because the article covers a historic act of violence. Documenting wars, terrorist attacks, and other such violent atrocities on Wikipedia is likely to raise tension among anyone who may have a personal connection to these events. For some annotators, it can be challenging to distinguish between a discourse that is an impassioned, albeit crass, plea for the accurate portrayal of violent conflicts and that which is a toxic discourse eroding Wikipedia's community norms around civility. 


Overall, this exploratory analysis reveals significant divergence in the type of language that each cluster of annotators perceive to be \textit{toxic} or \textit{severely toxic}. When we consider that machine learning models are typically trained from the mode of the annotation values for each comment, such divergent perspectives can become a clear source of unintended bias in a downstream model attempting to identify toxic language in online communities. When we think of crowd annotators as interchangeable units of human computation, it is easy to miss that the people who annotate such data contain a similar diversity of perspectives and opinions that occur in the online community whose data they are judging. Position mining allows us to infer these perspectives and relate them to the contexts which inspire the friction, i.e. toxicity, that automated content moderation algorithms serve to mitigate.

This analysis sets up a critical reflexive opportunity for us, as data scientists and researchers.
While we did not design the annotation task, we have an opportunity to reflect on whether our
position with regard to the research context is one that favors an online community in which
the tone of discourse is of equal or greater importance to a community member's freedom
to express themselves in an aggressive manner as long as the intention is to yield
a better article for the broader community. This is a central tension discovered
by \cite{duggan_online_2014, duggan_online_2017} in which their survey on online harassment 
suggests that $56\%$ of men believe it is more important to 
``Be able to speak their minds freely online'' than ``Feel welcome and safe online'' while women
many women preferred the latter ($63\%$). As the data scientists conducting this analysis, we
have a preference for online communities that favor making its users feel welcome and safe regardless
of the (likely minimal or non-existent) adverse effect this could have on the quality of the articles.
However, this reflexive insight is not necessarily easy to capture in the annotation task which is 
something we investigate in the next section.

\subsection{Locating the Data Scientist}
Our position mining results reveal that there are least two major groups of annotators based on their subjective labeling judgements. 
These two positions could be easily misrepresented by aggregating crowd labels into a singular perspective. While it's important for the data scientist to understand the various perspectives and positions of the annotators, reflexivity encourages us to acknowledge \textit{our place in the research} and understand ourselves relative to the annotators, the models being developed and deployed, and the
broader sociotechnical system. 
To facilitate this understanding---and as part of a broader computationally reflexive data science practice---we take a reflexive turn and create our own annotation fingerprint. As a practice, this allows us to position ourselves relative to the broader pool of annotators as well as reflect more deeply on the annotation task itself. In what follows, we present a first-person account of the first author using the techniques in this paper to create their own annotation fingerprint, understand their position relative to the broader pool of annotators, and discuss the reflexive insights that emerge from the processes involved in this practice. 

\subsubsection{A Self-Reflection: Creating My Annotator Fingerprint as a Data Scientist}
My goals in creating my own annotator fingerprint were to understand my annotation choices in relation to others, to understand my annotation choices in relation to modeling choices I might make, and to learn about the broader context of the annotation task. To understand
more about how annotators may be making decisions with regard to items where the task can feel ambiguous,
I used the following questions to guide my process:

\begin{itemize}
    \item \textit{Which positional cluster am I most aligned with?}
    \item \textit{What kind of ambiguity in the annotation task is being leveraged when annotating complicated comments?}
    \item \textit{What kind of ambiguity in the comments are being leveraged when choosing an annotation?}
    \item \textit{How do I reason in situations of ambiguity?}
    \item \textit{How could annotators be reasoning in situations of ambiguity?}
\end{itemize}

The first choice I needed to make in creating my annotator fingerprint is which data to annotate.
Instead of taking a random sample which would be likely to mostly yield neutral or uncontroversial
items, I decided to specifically sample items from varying levels of divisiveness between
the two positional clusters to get a sample that evenly represents the spectrum of content
that they agree on to the content they disagree on. To achieve this I grouped the data by
the difference in each positional cluster's modal label and sampled 13 comments from 
each group resulting in a total of 97 comments\footnote{One group had fewer than 13 comments to draw from.}. After shuffling these comments randomly, I began annotating the comments using the same prompt as the original annotators. 

\subsubsection{My Observations During the Annotation Process}
In addition to annotating the data to establish my own annotator fingerprint, I also recorded observations that might clarify some of the questions raised above. While my process did not strictly adhere to a formal process of reflexive journaling as a professional qualitative researcher might, it was a personally helpful process that provided context for some of the more difficult aspects of model building for the purpose of addressing larger systemic issues, like toxic behavior, in a large
complex sociotechnical system, like Wikipedia. Below are some of the highlights from the process:

\paragraph{\textbf{Some comments provide a decontextualized glimpse into an ongoing conflict among Wikipedia editors and that can be very hard to annotate.}} As the process of building and editing a Wikipedia article turns from democratic to adversarial, editors begin discussing the conflict among the editors in addition to the topic itself. Some of the comments describe an ongoing saga of toxic behaviors by other editors, but as annotators we only see one comment as a representation of a much bigger story. This puts the annotator in a difficult position where they need to choose between the editor who wrote the comment they are reading and the other editor they are in conflict with in order to decide whether the comment would be likely to make the annotator leave the discussion. If the annotator believes the editor who wrote the comment is correct or just in this conflict, then it would make sense that the comment is either neutral or that it would make them want to continue the discussion. If the annotator believes the editor is wrong, then they will likely feel that the community is worse off because of the comment. These situations add another layer of subjective context that can only be inferred or imagined in
order to annotate the data. Here is an example of one such comment:

\begin{quotation}
\textit{I have just pointed out that Milbourneone is guilty it would seem to WP:CIVIL against me too... for the following 1. Direct rudeness   * (a) Rudeness, insults, name-calling, gross profanity or indecent suggestions;   * (b) personal attacks, including racial, ethnic, sexual and religious slurs, and derogatory references to groups such as social classes or nationalities;   * (c) ill-considered accusations of impropriety;   * (d) belittling a fellow editor, including the use of judgmental edit summaries or talk-page posts (e.g. ``snipped rambling crap'', ``that is the stupidest thing I have ever seen'');I refer to refs of you stating letter written by myself was of such poor quality as to be hardly readable. So when considering pointing out WP:CIVIL to others please look at your own writings and realise that you are not exempt from the same criticism. With respect. TruthBomb\footnote{This is a pseudonym username we chose for anonymization} (talk)}
\end{quotation}

\paragraph{\textbf{Some comments suggest that there may be a conspiracy to control the narrative of the topic}} Among the comments in which the two clusters' modal labels diverged by more than 3 points, there were a few which alleged an organized effort to control the narrative of a Wikipedia article. One possible reason this becomes divisive is that one cluster may see the possibility of such a conspiracy as plausible. In such a case, this makes the interlocutor's decision to call out the conspiracy something that is healthy for the community. If one does not find the conspiracy to be plausible, then the allegation might appear harmful for the community, and thus the annotator may rate it as highly toxic. Again, this adds additional layers of reasoning to annotating the dataset that were not originally accounted for in the design of the annotation task and label descriptions.

\begin{quotation}
\textit{Can Someone Lock in the New Links that CIA Keeps Deleting? I have added what I consider, after 20 years of advocacy against fierce opposition from CIA and its FBIS minions,a few essential links.  I don't have the time or energy to fight the morons.  If there is an adult with Wiki authority to lock in the links I have added, similar to the manner in which the CIA links are locked in (I have more integrity than they do and would NEVER consider deleting their links), then I think we are all better off for.  If not, www.oss.net will remain up forever, and continues to be *the* reference site for OSINTneither the government nor the vendors are honest on this topic.}
\end{quotation}

\paragraph{\textbf{Some subjects are very difficult to talk about in a way that can be broadly perceived as civil or polite}} Some topics are notoriously difficult to discuss in a manner that is comfortable for everyone in Wikipedia's global audience. For example, some Wikipedia editors document recent international conflicts in which members of each side are motivated to assert their perspective. In other situations, Wikipedia editors may be discussing specific details about traumatic historical moments within an article's talk page and these editors are likely completely unaware that their comments will be judged, out of context, by crowd annotators who were not prepared to consider the challenges of documenting these events. In the example below, a Wikipedia editor discusses a cultural geographer who researched ``free people of color'' in colonial Virginia. Personally, I had a hard time annotating this comment, because I am aware that there is a great deal of ``research'' from this time, specifically Eugenics and Phrenology, which used academic authority and rhetoric to harm specific communities. Without the broader context of this discussion among editors or any personal familiarity with this very specific topic, I was not confident that I could make a correct or even benign decision. It is likely that many annotators were similarly unprepared for such situations:

\begin{quotation}
\textit{...He found that most could be traced to free mixed-race families formed of descendants of unions between white women and African men in colonial Virginia.[4] They migrated along with European American neighbors to frontier areas, where they found less restrictive racial conditions. In some cases, he found that descendants consistently ``married white``, and had children of increasingly European-American or white appearance.[...]We should only use stuff in a article about melungeons which is directly about Melungeons.  The person who added this attempted to use ``free person of color`` to mean african american when the law stated Free Person of color was not just african american, it was a catch all term.[..]Free person of color would mean any race that was not pure white.  Either way the paragraph should be removed as it does not mention Melungeons at all nor does the sources used.}
\end{quotation}

In performing my own annotation task to establish an annotator fingerprint, I initially expected to learn something about how to optimize the model. In actuality, I learned that ambiguity abounds and that many of my choices in the annotation task did not have an obvious connection to my values as I would have
described them prior to this exercise. Reviewing the annotations which represent the spectrum of disagreement between two major perspectives in this dataset helped me to appreciate the plurality of valid positions with regard to ``toxic comments''. While the examples above are just a few that arose, it became clear that this practice may help data scientists to develop a broader awareness of the limitations of an annotation task to enable annotators to express their personal values. Additionally, this reflexive practice can help 
data scientists to further consider the more complete social, cultural, and political provenance of their models 
and develop an understanding that they are embedded in a rich sociotechnical context that 
can reinforce, promote or suppress certain values. 

\subsubsection{Locating Myself With Respect to Crowd Annotators}
\begin{figure}
    \centering
    \includegraphics[scale=0.28]{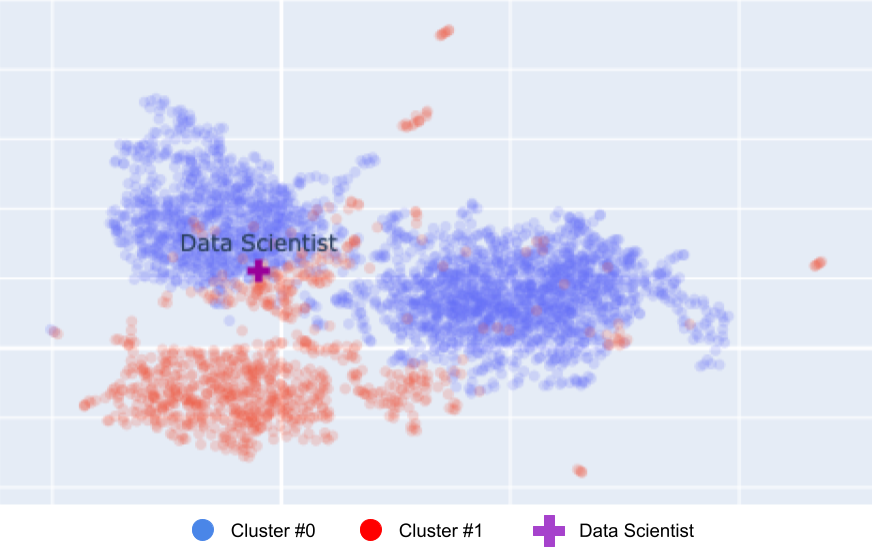}
    \caption{Annotator Fingerprints with Data Scientist's annotator fingerprint projected to a two-dimensional space using UMAP for dimensionality reduction.}
    \Description{An X-Y graph that shows a two blue clouds of data points in the middle and upper left representing cluster 0 and a red cloud in the lower left representing cluster 1. The data scientist is indicated by a plus at the right bottom edge of the upper left blue cloud.} 
    \label{fig:ann_fp_w_ds}
\end{figure}

Figure \ref{fig:ann_fp_w_ds} reveals that my personal fingerprint is situated in the majority Cluster $0$ but near its border with Cluster $1$. I was expecting to find myself located more central among the majority cluster $0$ given my observation that the members of Cluster $1$ tend to judge some of the comments in which one Wikipedia editor berates another with profane language as still being healthy for the community. When I noticed that my fingerprint was situated closer to the border with Cluster $1$ than expected, I recalled situations in which an editor was reiterating the profane language of another editor in an effort to highlight their lack of decorum. Upon further reflection, it became clear to me that there is an inherent difficulty in annotating a comment which references comments and behaviors by other editors, because the comments of the editor being referenced cannot be directly observed in the annotation task. The comment is both a call for the support of other members of the community to resolve the conflict by banning the editor that seemingly lacks decorum as well as a comment that uses the same profane language that expresses this lack of decorum. This nuance is important and yet it remains ambiguous to the annotators, to the data scientist, and inevitably to the model which will learn from this annotated dataset. By locating myself, as the data scientist responsible for creating a model and  automating a position, among general groups of annotator positions, I create an opportunity to reflect on a divisive aspect of annotating for a subjective concept and make salient the deliberate choices I made with regard to specific and important contexts in this online community.


\subsubsection{Locating the Model Among Human Annotator Fingerprints}
\label{comparing-model-fingerprints}

\begin{figure}
    \centering
	\includegraphics[scale=0.32]{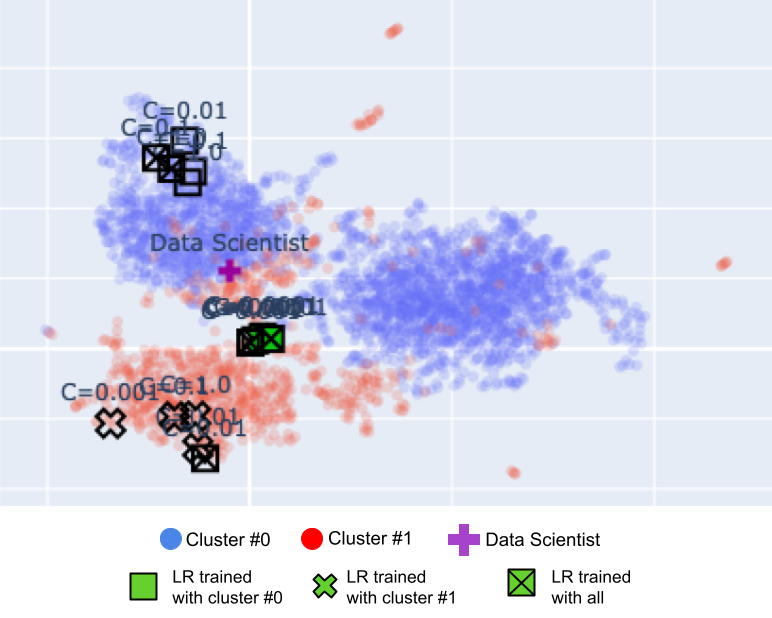}
	\caption{Map of Annotator Fingerprints with Model Fingerprints}    
	\Description{An X-Y graph showing the red and blue clusters, the data scientist and also the position of the various trained models with respect to the clusters.}
	\label{fig:fp_map_zoomed_out}
\end{figure}

\begin{figure}
	\centering 
	\includegraphics[scale=0.32]{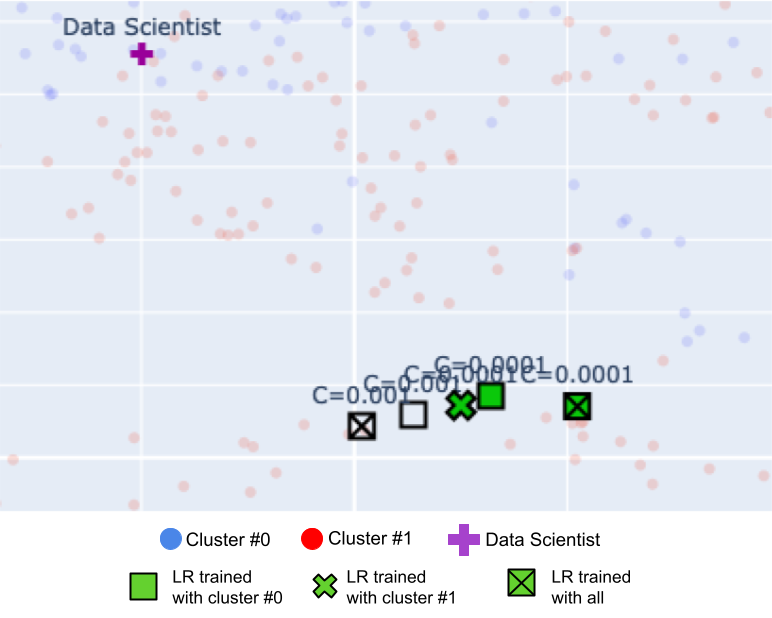}
	\caption{Zoomed to map area containing Data Scientist fingerprint and model fingerprints with optimal C parameters.}         
	\Description{A zoomed-in X-Y graph showing the red and blue clusters and a zoomed in look at the data scientist position and the trained model positions with respect to one another.}
	\label{fig:fp_map_zoomed_in}
\end{figure}

Figures \ref{fig:fp_map_zoomed_out} and \ref{fig:fp_map_zoomed_in} illustrate annotator fingerprints that include models trained solely on data with labels from Cluster $0$ represented as a square, models trained solely on data with labels from Cluster $1$ represented as an \textbf{X}, and models trained on data with labels from all of the crowd annotators represented as a square containing an \textbf{X}\footnote{These demonstration models are based on logistic regression models which use word count features as input.}. 

In reflecting on figures \ref{fig:fp_map_zoomed_out} and \ref{fig:fp_map_zoomed_in}, I am careful
not to over-interpret certain patterns within a two-dimensional projection of a high-dimensional dataset.
This is because the dimensionality reduction process is stochastic and generally lacks a reliable 
validation process that is robust to different data structures. Each pattern observed, should 
serve as a curiosity that may be worth exploring through other means. One of the primary goals of
computational reflexivity is to offer an ecosystem of complementary methods that allows
the data scientist to address the limitations of one method using another similar to 
many-model thinking \cite{page_why_2018}.

Some of the patterns observed align with general expectations. For example, models trained
on a specific cluster's labels, tend to result in model fingerprints that are most aligned
with that cluster. However, clusters trained with all of the data, but with different 
hyperparameter values can vary dramatically in which cluster they align with and how strongly.
Figure \ref{fig:fp_map_zoomed_out} shows the classifier trained using different
regularization parameter values for $C$. This parameter is intended to mitigate overfitting 
by reducing variance. Smaller values yield stronger regularization which yielded the models
which were more optimally able to predict the modal value of all labels (depicted by the green \textbf{X}'s and squares). As the value for $C$ increases and thus the regularization becomes weaker
and allows for more variance, the position that the model aligns with also varies fairly dramatically.
This type of observation helped me to see the bias-variance trade off in terms of model's
alignment with multiple stakeholder perspectives, instead of only considering validity
in terms of its alignment with a single set of labels that represents a reduction of this plurality
through aggregation.

Another observation worth noting is that the models which would have been selected 
through a traditional validation process (e.g., best RMSE scores compared to the mode of all labels for each item)
are positioned within an area with some of the lowest density of annotator fingerprints. This could
be evidence supporting my hypothesis that using these aggregated labels creates a perspective 
that is not well aligned with anyone. This is akin to a common problem in statistics when 
an analyst might assume that the mode of a distribution is the best representation of the data, while
overlooking the fact the data might be multimodal. If it is the case that the aggregate label itself
can't be trusted to adequately represent the annotators, because the annotators are generally falling into
two distinct positions, then it may be worth making a value decision about which position
is most aligned with the overall mission of the organization. Reflexivity in data science can help us to
understand ourselves in the context of the data, the model, and the broader research, but to also understand when a decision or model should be made through reflection of one's own values as much as it is made through data.

\subsection{Concluding Thoughts on the Case Study}
Through this case, we provided a demonstration of how we can use the annotation fingerprinting process to locate the data scientist and their models with respect to the larger crowd used to annotate the training data. Processes like these could be further developed to facilitate greater computational reflexivity for the field of data science and provide critical context for the pattern of decisions a model makes by relating these decisions to those of the human crowd annotators and data scientist. Similar to the benefits of practicing reflexivity in qualitative research, computational reflexivity -- and tools to support its practice -- could help data scientists develop broader awareness of their models as embedded in rich sociotechnical systems and understand the impact this may have on promoting or suppressing certain values of an online community. Combining such tools with common model explanation techniques \cite{markham_2017, lundberg_unified_2017} can further enable the data scientist to contextualize a learning algorithm's inductive bias. Furthermore, such a reflexive mode of analysis gives the data scientist the tools and data needed to have a conversation with the rest of their organization about which values are being promoted. For example, a data scientist equipped with a journal of qualitative observations and hypotheses from the process of producing their own annotation fingerprint like those discussed here, might organize a discussion about what values should be promoted and which should be hampered through the automated content moderation model being developed.

\section{Discussion}
Data scientists are often charged with revealing novel insights from
large datasets that can ultimately serve as the basis for life-altering decisions such as who will receive parole or who will be among the first to receive a 
COVID-19 vaccine \cite{green_good_2019, wiggers_covid-19_2020}. In such situations, a great deal of trust and power is placed in data scientists -- and this dynamic should be critically examined. As others have observed, the discretionary decisions that data scientists make can result in significant downstream consequences \cite{10.1145/3411764.3445518}. While many data scientists would like to proactively mitigate these consequences \cite{10.1145/3411764.3445518}, they often need to negotiate critical aspects of their work with other stakeholders \cite{10.1145/3411764.3445518, passi_problem_2019} and wait for the consequences to have an adverse and measurable effect on end-users before they can begin to resolve the issue \cite{10.1145/3290605.3300760}. 

Fortunately, there has been a lot of recent momentum to address these challenges in various academic communities. Critical data studies has offered an important perspective on the challenges of having ``big data'' as a central tenet of modern society \cite{boyd_critical_2012, iliadis_critical_2016, dalton2014does}. Human-Centered Machine Learning \cite{gillies_human-centred_2016, fiebrink_introduction_2018, guestrin_4_2019, cambo_user-centred_2018} and Human-Centered Data Science \cite{10.1145/3323994.3369898, aragon_developing_2016, 10.1145/3290607.3299018, 10.1145/3406865.3418584} are two communities that have been discussing and studying the connection between data science work and the impact it has on all relevant stakeholders. The FAccT community spans law, policy, computer science, education, sociology, and philosophy to progress our understanding of how we can develop sociotechnical systems with a core focus on fairness, accountability, and transparency \cite{mitchell_model_2019, chancellor_taxonomy_2019, bender_stochastic_parrots_21, 10.1145/3442188.3445888}.
Together these communities are developing theories, methods and systems that better understand and mitigate the adverse effects of automated decision-making tasks performed by machine learning systems. As others have argued \cite{elish_situating_2018, 10.1145/3442188.3445880, Tanweer2021Why, dignazio_data_2020}, the concepts of reflexivity and positionality can provide a framework for data scientists to more systematically address and communicate the influence of discretionary decisions in the development of machine learning models. Put another way, reflexivity and positionality help data scientists to contextualize and situate their own knowledge, the knowledge held by data annotators, and the knowledge represented by computational models in a scalable, intuitive, and visual manner.

To understand what computational reflexivity might look like in practice, consider the challenge of stakeholder (mis)alignment in data-driven organizations \cite{passi_problem_2019, passi_trust_2018, 10.1145/3442188.3445880, 10.1145/3411764.3445518}. Stakeholders in the development and application of machine 
learning can include the data scientist, the annotators, and the users as well as those in the larger business context like project managers, product designers, and business analysts \cite{passi_trust_2018}.
Conceptual and epistemic misalignment among these stakeholders can affect problem formation \cite{passi_problem_2019}, data and model validation metrics, methods, and criteria \cite{passi_trust_2018}, and the ability of a data science team to implement practices of fairness \cite{10.1145/3411764.3445518}, accountability \cite{10.1145/3442188.3445880}, and transparency \cite{10.1145/3442188.3445880}. Many have argued that adopting reflexive practices can help data scientists and 
data-driven organizations to understand the impact of each stakeholder's position in the production of knowledge \cite{elish_situating_2018, Tanweer2021Why, 10.1145/3442188.3445880}, and these methods can help data scientists to reflect on their conceptualization of the problem and intuitions for how to solve it. When the product of these exercises are shared and communicated with other stakeholders, the best case scenario is that it presents opportunities for alignment. In the worst
case scenario, it presents opportunities to create documentation 
that improves transparency, accountability, and auditing as Miceli and colleagues suggest \cite{10.1145/3442188.3445880}.

While existing approaches provide important components for addressing fairness, accountability and transparency, they are still somewhat limited in that they are grounded in each stakeholder's own thinking. For example, while the reflexive practices suggested by Tanweer and colleagues \cite{Tanweer2021Why} are crucially important for data science, they primarily involve self-reflection. We can augment the reflexive practices they recommend by incorporating
computational reflexivity techniques to broaden the context of the 
reflexive work so that it includes the data and the model. 
Computational reflexivity aims to answer the call for computational methods that account for the nuanced and situated nature of knowledge production \cite{10.1145/3323994.3369898} by providing techniques to better reflect on our position with regard to many data points, many annotators, or many model predictions and exposing their relevant context. While self-reflection is critical for providing the context necessary to situate knowledge, Miceli and colleagues \cite{10.1145/3442188.3445880} argue that a \textit{relational} examination \textit{between stakeholders} can further reveal important social context such as power dynamics that enable stakeholders in vulnerable positions to raise questions that wouldn't have been considered otherwise. Computational reflexivity enables this relational examination to occur in contexts like the Wikipedia toxic comment challenge dataset where there are many annotators (3,591), many data points (159,686) and many annotations (1,598,289). In the following section, we discuss how computational reflexivity addresses the issue of scaling reflexive practices to meet the needs of this relational examination.

\subsection{Practicing Computational Reflexivity in Data Science}
The central goal of reflexivity is to contextualize each stakeholder's self within the research context to enable consideration of the influence that social and epistemic positions have in knowledge production. When applying reflexive practices in the context of ``big data'' research, the issue of scale can become hard to ignore given the volume of data, the number of crowd annotators, the number of annotations they produce, and the shear scale of automated decisions that are made by a machine learning model in both a research and development context (training and validation) and the context of real-world application (deployment). Computational reflexivity simultaneously offers a solution to the challenge of scale and presents new opportunities for reflexive thinking. For example, we may solve the issue of scale with regard to the volume of data through data reduction as we did with topic modeling in section \ref{sec:case_study}, but 
how do we know that the ``topics'', as defined by the topic modeling algorithm, are the appropriate representation in this research context? Data scientists may intuitively approach these questions from a statistical and computational perspective by considering the topic model's reconstruction error
as we did in \cite{cambo_2020}. However, the choices about what information is represented by a topic and what information is lost are worth reflecting on -- not to prove that we made the best decision, but to show that we made a good decision and that we understand that it was a discretionary decision and to openly acknowledge our role and its impact on the final outcome. 
Thus, the goal of computational reflexivity is not necessarily to prescribe answers to important technical questions, but rather to strengthen the decisions we make in answering these technical questions by connecting them
with related reflexive questions. 

Understanding one's self in the context of one's own research can further benefit from understanding one's self in the context of other stakeholders in the research. For example, data scientists can ask others involved in the development process to annotate enough data for annotator fingerprints to be made and compared.
An internal development team (e.g. project manager, product designer, software engineer) might consider highlighting variances at each cell within the annotator fingerprint matrix as a discussion prompt that can aid the development process by drawing attention to conceptual misalignments with regard to the data. However, when we consider machine learning systems that utilize 
models trained on datasets labeled through a crowd annotation process, the crowd annotators are 
perhaps the most consequential influence on the system's behavior. The most common way that
annotators are considered in the crowd annotation process has been through the use of 
inter-rater reliability methods like Cohen's Kappa \cite{cohen_statistical_1988} or 
Krippendorff's Alpha \cite{krippendorff_computing_2011}. However, such measures are extremely reductive in the way that they aggregate the many determinations made by many annotators in the many varying contexts of the data. Our annotator 
fingerprinting and position mining approach, as well
as the CrowdTruth techniques proposed by \cite{dumitrache_crowdtruth_2018, mika_crowdtruth_2014}, offer a higher fidelity, yet scalable approach to achieving a better understanding of how annotators 
approach such tasks. These techniques give data scientists insight into the broader social context which may lead to the identification of certain 
positional biases that could be harmful. Returning to our case study, applying a computationally reflexive approach highlighted the possibility that there are two major perspectives at play in the crowd annotation task: the perspective that civility is greater than the pursuit of article quality, and the perspective that article quality is more important than civil discourse. These two major perspectives help to take general questions that may come up in reflexive practices like \textit{``What do I think is most important for users of this platform?''} and \textit{``What do I think discourse on this platform should be like?''} and make them more specific and contextualized like \textit{``Do I believe that civil discourse is more important than article quality?''} or \textit{``Does the organization value civil discourse more than article quality?''}. Consideration of and alignment among these questions may lead to new design considerations for the machine learning system being developed.

While datasets and their annotations are heavily influential to a model's final inference behaviors, each machine learning algorithm presents its own inductive biases based on assumptions about how data relates to the real world. It can be very difficult to foresee how the inductive biases of each algorithm will play out in the context of each new dataset. Model explanation techniques like LIME \cite{ribeiro2016should} can help data scientists to better understand the mechanics of a model's inference process, which can be useful for understanding inductive biases, but human social context is most necessary for understanding
whose perspective this model best represents and whose perspective it least represents. Binns et al. \cite{binns_like_2017} have studied the
connection between annotator groups and model predictions by manipulating the proportion of annotators by gender in the annotation data. What they found was that inductive biases in the algorithm
can favor one group over another even if the model was trained exclusively with annotations from the latter group. For example, if one group of annotators typically base their annotation decisions on the presence of specific words that trigger a specific reaction, while another group engages more deeply with what the text implies, then it is reasonable to believe that most learning algorithm will favor the former group's perspective even when trained completely from data in the latter group, because of machine learning's inability to thoroughly reason
about the data from a human perspective. By framing model behaviors as model positionality and comparing model behaviors to annotator behaviors in a similar task and context, we can better understand who benefits from this process and who is disadvantaged by it.

\subsection{General Considerations}
In this section, we present some additional ideas to consider for more general work in data
science methodology and pedagogy.

\textbf{Pragmatic critiques can appeal to those struggling with the idea that data is a sociotechnical concern.} Data science has an image of being an exclusively technical discipline which can make it difficult for data scientists to engage with sociocultural critiques \cite{bates_integrating_2020}. Some researchers have observed that even when data scientists are interested and willing to address social concerns, they don't always have the expertise or organizational support to do so \cite{10.1145/3411764.3445518}. However, such critiques can sometimes be reframed as practical challenges. For example, a critical theorist might present the idea that data encodes and exacerbates existing power differentials and inequalities. This statement is well supported by evidence and often resonates with those familiar with critical theory. However, such a statement often relies on prior knowledge to understand how data and social inequality relate. It may also alienate students coming from positivist fields such as physics or economics that may not share the understanding that technology is often wielded as a tool for cultural imperialism. By complementing the more constructivist ideas from critical studies with the practical challenge they create, we may be able to bring more students, scholars, and practitioners into the conversation. We can take a step back and ask data scientists to consider the challenge of creating a single, accurate, and generalizeable model of a complex social phenomenon which is understood differently in different contexts by different communities. 

Computational reflexivity helps to bridge the gap between commonly used methods in data science which focus on overcoming practical challenges with ``big data'' and reflexive practices in qualitative research which focus on overcoming 
the practical challenges of mitigating personal bias and fully recognizing our subjective interpretations of empirical observations. For example, inter-annotator agreement, 
a measure that most data scientists are familiar with, can be extended to annotator fingerprinting and CrowdTruth methods which can be used to not only assess the quality
of a labeled dataset, but also to explore the landscape of subjective interpretations embedded within the labeled dataset. Similarly, position mining can aid a data scientist's understanding of the critical points made by scholars in critical data studies by contextualizing an important concept from modern qualitative research, positionality, within a familiar computational method, clustering.


\textbf{Data can be biased in ways that we didn't know to look for.} A short time ago, most papers which studied algorithmic bias did so through the lens of either gender or race, in part, because these were aspects of identity which have been systematically oppressed in ways that are salient in fields like sociology and psychology. 
However, while we have theoretical guidance and methods that can help us look for biases that we were previously aware of, we have a less developed toolkit for identifying potential biases that we haven't considered looking for. The position mining methods detailed in this work aim to group annotators by their common behavioral biases exhibited via their annotation patterns. Such an approach can help to identify biases that we did not, a priori, know to look for as well as reveal those that we did not think to collect data to examine. For example, the dataset explored in this paper includes annotator age, education, language, and gender, but it does not include race, sexual orientation, or political orientation, each of which have been identified as the most common aspects of identity that were targeted in online harassment \cite{duggan_online_2017}. By identifying groups of 
annotators who have common annotation biases and are distinct from other groups of annotators, we may be able to identify biases that were not obvious at the time the ontology was constructed or the data was annotated and collected.

\textbf{Data size is a non-negligible issue for modeling.} Many data science techniques are criticized for being overly reductionist. For example, a common approach to understanding whether men and women annotate toxic language differently is to compare the mean and standard error of Likert-scale annotations for each group. Such an approach posits that gender is 1) a discrete and binary trait, and 2) a meaningful grouping mechanism. This ignores that adding the dimension of race would highlight differences where gender alone might not. That said, the interaction of race and gender may also be insufficient, and we can keep adding more facets of identity until each group is determined to be sufficiently homogeneous in terms of their demographic traits. The resulting groups may be too small for either research or model production. Annotator fingerprinting offers an alternative by using annotation data with common clustering algorithms to identify groups defined by their \textit{common annotation behaviors} and not their \textit{common facets of identity}. This permits groupings which may be large enough to have a major impact on the perspective encoded through the annotation aggregation process and which are large enough to build reasonable machine learning models from.

\section{Limitations and Potential Misuse}
When researchers reflect on their place in the work, their epistemological framing, 
and their methodological choices, it is important that they consider the implications of having made different choices. Similar to Gelman's 
``garden of forking paths'' \cite{gelman_garden_2013}, it can help to consider the implications of having made a different, but similarly valid choice in the research process. Position mining and computational reflexivity can help guide the data scientist to consider plausible alternatives for the annotation set, 
data processing techniques, learning algorithms, and model selection criteria. 
However, it may have little effect for those in machine learning research who choose evaluation techniques which best support a claim, novel finding or performance gain. It is also possible that the complexity added by the techniques presented in this paper could be used to obfuscate or mediate 
results that do not support an enthusiastically positive narrative around a novel technical contribution in research.

The audience that reads this work will similarly need to adjust their understanding of how evidence supports a claim. Traditionally, machine learning research publications have accepted claims that a learning algorithm can generalize to new data if the research presents evidence that the learning algorithm can produce a classifier that can accurately predict on held out data from the same dataset as the training data. However, the work we presented on the toxic comment classification dataset suggests that data annotated by a heterogeneous group of crowd workers and aggregated into a single label reduces the complexity of a sociotechnical phenomenon produced by a diverse collection of users into a model of a monoculture which simply does not exist. However, with public discussion around high profile cases of failed data science projects by large, well-resourced companies like Facebook, Google, YouTube, and others, we expect that the audience for such research will be more easily convinced that evaluating algorithms requires more critical reflection. In fact, recent work has shown how qualitative and machine learning methodologies compare, contrast, and combine to create new analytic processes \cite{muller_machine_2016, baumer2017comparing}.

\section{Conclusion}
Data science is an inherently eclectic field that was initially conceived as the intersection of math, computer science, and subject expertise. While many have tried to reduce data science to an engineering discipline, we must remember that data is born from all aspects of the world and its pedagogy and epistemology should reflect that. For problems and domains that involve the social, the cultural and the political our tools and approaches must bring a greater appreciation for the ways in which complex social phenomena operate in the context of large-scale, global, sociotechnical systems. It is imperative that the coming generation of data science researchers and scholars commit not only to the technical components of the discipline, but also to a responsibility to engage, understand and involve those with more direct experiences in the subject domains in which they work. It is our hope that computational reflexivity can serve to provide practitioners and researchers alike with the tools and understanding to make this vision a reality.

\begin{acks}
This work was supported, in part, by a graduate research grant from Northwestern University. We thank Anne Marie Piper, Bryan Pardo, Nick Diakopoulos, Mark Díaz and members of the CollabLab for the indispensible feedback that has helped us to make this research better. We also thank the anonymous reviewers for their valuable feedback.
\end{acks}

\bibliographystyle{ACM-Reference-Format}
\bibliography{base,new}


\end{document}